\documentclass{ws-ijmpa}

\newcommand{\ii}{\mathrm{i}}
\newcommand{\dd}{\mathrm{d}}
\newcommand{\ee}{\mathrm{e}}

\begin{document}

\markboth{A.\,V.~Kuznetsov, A.\,A.~Okrugin}
{The Exact Electron Propagator in a~Magnetic Field as the Sum\dots
}

%
\catchline{}{}{}{}{}
%

\title{THE EXACT ELECTRON PROPAGATOR IN A~MAGNETIC FIELD \\
AS THE SUM OVER LANDAU LEVELS \\
ON A BASIS OF THE DIRAC EQUATION EXACT SOLUTIONS
}

\author{A.\,V.~KUZNETSOV and A.\,A.~OKRUGIN}

\address{Division of Theoretical Physics, Department of Physics,\\
Yaroslavl State P.\,G.~Demidov University, Sovietskaya 14,\\
150000 Yaroslavl, Russian Federation, Russia\\
avkuzn@uniyar.ac.ru, okrugin@uniyar.ac.ru}

\maketitle

\begin{history}
\received{Day Month Year}
\revised{Day Month Year}
\end{history}

\begin{abstract}
The exact propagator for an~electron in a~constant uniform magnetic field as the~sum over Landau levels is obtained by the direct derivation by standard methods of quantum field theory from exact solutions of the Dirac equation in the magnetic field. The result can be useful for further development of the calculation technique of quantum processes in an external active medium, particularly in the conditions of moderately large field strengths when it is insufficient to take into account only the ground Landau level contribution.

\end{abstract}

\ccode{PACS numbers: 14.60.Cd, 12.20.Ds, 02.30.Gp}

\section{Introduction}
\label{sec:Introduction}

Among astrophysical objects there exists a~class of neutron stars which are called magnetars. Magnetic field values exceed there the critical value of $B_e = m_e^2/ e \simeq 4.41 \times 10^{13}$~G, where $m_e$ is the electron mass.\footnote{We use the Planck units: $\hbar = 1$, $c = 1$, and the Minkowski metric with the signature $(+, -, -, -)\,$.}
There is no longer possible to take into account the field influence by the perturbations theory under such conditions in the analysis of quantum processes. The detailed description of the calculation technique of processes in external fields one can find, for example, in reviews and books, see Refs.~\refcite{Ritus:1979}--\refcite{Kuznetsov:2003}.

The magnetic field influence on the particle properties is determined by the specific charge, i.\,e. by the particle charge and mass ratio. Hence, the charged fermion which is the most sensitive to the external field influence is the  electron. The calculations of specific physical phenomena in strong external field are based on the application of Feynman diagram technique generalization. It consists in the following procedure: in initial and final states the electron is described by the exact solution of the Dirac equation in the external field, and internal electron lines in quantum processes correspond to exact propagators that are constructed on the basis of these solutions.

The expression for the exact electron propagator in the constant uniform magnetic field was obtained by J.~Schwinger\cite{Schwinger:1951} in the Fock proper-time formalism.\cite{Fock:1937} There are another propagator representations given in a~number of works. Thus, in Ref.~\refcite{Loskutov:1976} the case was considered of superstrong field and the contribution of the ground Landau level to the electron propagator was obtained.
In Ref.~\refcite{Chodos:1990}, see also Ref.~\refcite{Chyi:2000}, the propagator was transformed from the form of Ref.~\refcite{Schwinger:1951} into the sum over Landau levels. Also in Ref.~\refcite{Chyi:2000} the electron propagator decomposition over the power series of the magnetic field strength was given .

In our opinion, it is quite important to know different representations of the electron propagator in the external magnetic field and the conditions of their applicability. There were some examples where misunderstanding of such conditions has led to erroneous papers. Thus, in Refs.~\refcite{Elizalde:2002,Elizalde:2004} the self-energy operator of neutrino in the magnetic field was calculated by the analysis of the one-loop diagram $\nu \to e^- \, W^+ \to \nu$. The authors of the paper restricted themselves by consideration of the ground Landau level contribution to the electron propagator. As was shown in Ref.~\refcite{Kuznetsov:2006}, the ground Landau level contribution is not dominant and the  next levels give the contributions of the same order of magnitude.

As we know, there is no such methodologically important issue in the literature as a direct derivation by the standard quantum field theory methods of the exact electron propagator in the external magnetic field in the form of the sum over Landau levels from the exact solutions of the Dirac equation in a magnetic field. The present paper is intended to fill this gap. 
The exact solution of the Dirac equation for an electron in the external magnetic field on the $n$th Landau level is given in Sec.~\ref{sec:Dirac-eq-solution}. On the basis of these solutions, the detailed derivation of the electron propagator is performed in Sec.~\ref{sec:propagator-calculation}. As a result, the propagator is written in $x\mbox{-}$representation as the~sum over Landau levels. In Sec.~\ref{sec:electron-propagator-as-sum} the identity is shown of the obtained expression for the propagator to the known result.\cite{Chodos:1990}

\section{The Solution of the Dirac Equation for an~Electron in the Magnetic Field}
\label{sec:Dirac-eq-solution}
%
The Dirac equation for an electron with the mass $m$ and the charge $(- e)$, where $e > 0$ is the elemetary charge, in magnetic field $B$ with the $4\mbox{-}$potential $A^\mu (X)$, where $X$ denotes the $4\mbox{-}$vector, $X^\mu = (t,x,y,z)$, takes the form:
\begin{equation}
\bigg(\ii \, (\partial \, \gamma) + e \, (A \, \gamma) - m \bigg) \, \varPsi(X) = 0 \,,
\label{eq:Dirac.Eqn}
\end{equation}
where $(\partial \, \gamma) = \partial_\mu \, \gamma^\mu$, $(A \, \gamma) = A_\mu \, \gamma^\mu$. For the case of the constant uniform magnetic field directed along the $z$-axis, choosing the $4\mbox{-}$potential as $A^\mu = (0, 0, x \, B, 0)$, it is possible to write down the so-called solutions with positive and negative energy.

The solution with positive energy is:
\begin{equation}
\varPsi^{(+)}_{n, \, p_y, \, p_z, \, s} (X) 
= \frac{\ee^{-\,\ii \, (E_n \, t - p_y \, y - p_z \, z)}}
{\sqrt{2\,E_n\,(E_n + m)\,L_y\,L_z}} \; 
U^{(+)}_{n, \, p_y, \, p_z, \, s} (\xi^{(+)}) \, ,
\label{eq:Dirac.Eqn.solution}
\end{equation}
where $n$ indicates the Landau levels: $n = 0, 1, 2, \dots$; 
$p_y$, $p_z$ are the electron ``momentum components'' along $y$ and $z$ axes (see below);
$L_y$, $L_z$ are the normalization sizes along $y$ and $z$ axes;
$s = \pm \, 1$ is the quantum number related to spin; 
$E_n$ is the electron energy on the~$n$th Landau level: 
\begin{equation}
E_n = \sqrt{m^2 + p_z^2 + 2 \, \beta \, n} \,, \qquad \beta \equiv e \, B \, .
\label{eq:E_n}
\end{equation}
The bispinor $U^{(+)}$ has different forms for the cases $s = - \, 1$ and $s = + \, 1$:
\begin{equation}
U^{(+)}_{n, \, p_y, \, p_z, \, s = -1} (\xi^{(+)}) =
\begin{pmatrix}
0 \\ \\
(E_n + m)\,V_n(\xi^{(+)}) \\ \\
-\, \ii \, \sqrt{2 \, \beta \, n}\,V_{n-1}(\xi^{(+)}) \\ \\
- \, p_z \, V_n(\xi^{(+)})
\end{pmatrix},
\label{eq:bispinor-1}
\end{equation}
\begin{equation}
U^{(+)}_{n, \, p_y, \, p_z, \, s = +1} (\xi^{(+)}) =
\begin{pmatrix}
(E_n + m)\,V_{n-1}(\xi^{(+)}) \\ \\
0 \\ \\
p_z \, V_{n-1}(\xi^{(+)}) \\ \\
\ii \, \sqrt{2 \, \beta\, n}\,V_n(\xi^{(+)})
\end{pmatrix},
\label{eq:bispinor+1}
\end{equation}
and for the ground Landau level, $n = 0$, the solution exists just at $s = -1$. The variable $\xi^{(+)}$ is related with $x$ coordinate by the relation:
\begin{equation}
\xi^{(+)} = \sqrt{\beta} \, \left(x + \frac{p_y}{\beta}\right).
\label{eq:xi}
\end{equation}
$V_n(\xi)$ are the harmonic oscillator functions expressed in terms of the 
Hermite polynomials $H_n(\xi)$:
\begin{eqnarray}
V_n(\xi) &=& \frac{\beta^{1/4}}{\sqrt{2^n \, n! \, \sqrt{\pi}}}\,
\ee^{-\,\xi^2 /2} \, H_n(\xi) \,, \qquad H_n(\xi) = 
(-1)^n \, \ee^{\xi^2} \, \frac{\dd^n}{\dd\xi^n} \, 
\ee^{-\,\xi^2} \, ,
\nonumber\\
&& \int_{-\infty}^{+\infty}|V_n(\xi)|^2\, \dd x = 1.
\label{eq:V_n}
\end{eqnarray}
It should be noted that in the above expressions $p_z$ is the conserved momentum component of the electron along the $z$ axis, i.\,e. along the field direction, while $p_y$ is the generalized momentum, which determines the position of the center $x_0$ of the Gaussian packet along the $x$ axis, $x_0 = - p_y/\beta$, see (\ref{eq:xi}). 

The solution $\varPsi^{(-)}$ with negative energy may be obtained from expressions~\eqref{eq:Dirac.Eqn.solution}, \eqref{eq:bispinor-1}\,--\,\eqref{eq:xi} by changing the sign of values $E_n$, $p_y$, $p_z$. The procedure of obtaining the solutions can be found, for example, in Ref.~\refcite{Kuznetsov:2003}.

\section{The Propagator Calculation on the Basis\\ of the Dirac Equation Solutions}
\label{sec:propagator-calculation}

To calculate the electron propagator, the standard method is applied based on using the field operators 
which include the Dirac equation solutions in a magnetic field:
\begin{equation}
\Psi(X) = \sum\limits_{n, \, p_y, \, p_z, \, s}
\Bigl(a_{n, \, p_y, \, p_z, \, s}\, \varPsi^{(+)}_{n, \, p_y, \, p_z, \, s}(X) +
b_{n, \, p_y, \, p_z, \, s}^{\, \dagger} \, 
\varPsi^{(-)}_{n, \, p_y, \, p_z, \, s} (X)\Bigr) \, .
\label{eq:field.operator}
\end{equation}
Here $a$ is the destruction operator of the electron, $b^{\,\dagger}$ is the creation operator of the positron, $\varPsi^{(+)}$ and $\varPsi^{(-)}$ are the normalized solutions of the Dirac equation~\eqref{eq:Dirac.Eqn} in a magnetic field with positive and negative energy correspondingly.

The propagator is defined as the difference of time-ordered and normal-ordered productions of the field operators~\eqref{eq:field.operator}:
\begin{equation}
S(X, X^{\,\prime}) = T \left( \Psi(X) \, 
\overline{\Psi}(X^{\,\prime})\right) -
{\cal N} \left(\Psi(X) \, \overline{\Psi}(X^{\,\prime})\right).
\label{eq:propagator_def}
\end{equation}
Using anticommutation relations for the creation and destruction operators, we obtain, that the propagator at $t > \, t^{\,\prime}$ and at $t < \, t^{\,\prime}$ is expressed in terms of the solutions with positive energy~\eqref{eq:Dirac.Eqn.solution} and negative energy correspondingly:
\begin{equation}
S(X, X^{\,\prime})\Bigr|_{t \gtrless \, t^{\,\prime}} = 
\pm \sum\limits_{n, \, p_y, \, p_z, \, s} 
\varPsi_{n, \, p_y, \, p_z, \, s}^{(\pm)}(X) \; 
\overline{\varPsi}\vphantom{\varPsi}_{n, \, p_y, \, p_z, \, s}^{\,(\pm)}
(X^{\,\prime})\,.
\label{eq:propagator_sum}
\end{equation}
Thus, the propagator is divided into the sum over Landau levels:
\begin{equation}
S(X, X^{\,\prime}) = \sum\limits_{n=0}^{\infty} \; S_n (X, X^{\,\prime}) \,.
\label{eq:propagator_sum_n}
\end{equation}
Further we will find the $n$th Landau level contribution into the propagator~\eqref{eq:propagator_sum}. It is convenient to come from the summation over the momenta $p_y$ and $p_z$ to the integration, by the substitution
\begin{equation}
\frac{1}{L_y \, L_z}\,\sum\limits_{p_y,\,p_z} \rightarrow 
\int \frac{\dd p_y \, \dd p_z}{(2 \pi)^2}\,.
\label{eq:sum2int}
\end{equation}
For the $n$th level contribution we found:
\begin{eqnarray}
S_n (X, X^{\,\prime})\Bigr|_{t \gtrless \, t^{\,\prime}} &=& 
\frac{1}{2\,(\pm E_n)\,(\pm E_n + m)}\,
\int\frac{\dd p_y \, \dd p_z}{(2 \pi)^2} \, \times
\nonumber\\[3mm]
&\times&
\exp\left\{\ii \left[\mp E_n(t-t^{\,\prime}) \pm  p_y (y-y^{\,\prime}) 
\pm  p_z(z - z^{\,\prime})\right]
\right\}\times
\nonumber\\[3mm]
&\times&
\sum\limits_{s = \pm 1} U^{(\pm)}_{n, \, p_y, \, p_z, \, s} (\xi^{(\pm)})\; 
\overline{U}\vphantom{U}^{\,(\pm)}_{n, \, p_y, \, p_z, \, s} 
(\xi^{(\pm) \,\prime})\, .
\label{eq:prop_n}
\end{eqnarray}
After a simple but quite cumbersome transformations one can reduce the matrices in Eq.~\eqref{eq:prop_n}, which are constructed from the bispinors~\eqref{eq:bispinor-1}, \eqref{eq:bispinor+1} and the corresponding bispinors of the solution with negative energy, to:
\begin{eqnarray}
&& \frac{1}{\pm E_n + m} \, 
\sum\limits_{s = \pm 1} U^{(\pm)}_{n, \, p_y, \, p_z, \, s} (\xi^{(\pm)})\; 
\overline{U}\vphantom{U}^{\,(\pm)}_{n, \, p_y, \, p_z, \, s} 
(\xi^{(\pm) \,\prime}) =
\nonumber\\[3mm]
&& = 
\frac{1}{2^n \, n!} \, \sqrt{\frac{\beta}{\pi}} \, \exp 
\left[-\,\frac{1}{2} \, (\xi^{(\pm)})^2 
-\,\frac{1}{2} \, (\xi^{(\pm) \,\prime})^2\right]
\bigg\{ 
\left( \pm E_n \gamma_0 \mp p_z \gamma^3 + m \right)\times 
\nonumber\\[3mm]
&&\times  
\left[ \varPi_- \, H_n (\xi^{(\pm)}) \, H_n (\xi^{(\pm) \,\prime}) 
+ \varPi_+ \, 2 n \, H_{n-1} (\xi^{(\pm)}) \, H_{n-1} (\xi^{(\pm) \,\prime}) 
\right] + 
\nonumber\\[3mm] 
&& + 
\ii \, 2 n \, \sqrt{\beta} \, \gamma^1 
\left[ \varPi_- \, H_{n-1} (\xi^{(\pm)}) \, H_n (\xi^{(\pm) \,\prime}) 
- \varPi_+ \, H_n (\xi^{(\pm)}) \, H_{n-1} (\xi^{(\pm) \,\prime}) 
\right]
\bigg\} \, ,
\label{eq:preob}
\end{eqnarray}
where the projection operators $\varPi_{\pm}$ are introduced:
\begin{equation}
\varPi_{\pm} = \frac{1}{2}\,(I \pm \ii \, \gamma^1\,\gamma^2) \,, \quad
\varPi_{\pm} \, \varPi_{\pm} = \varPi_{\pm} \,, \quad
\varPi_{\pm} \, \varPi_{\mp} = 0.
\label{eq:varPi}
\end{equation}
One can see, that after changing the signs of integration variables $p_y \to - p_y$ and $p_z \to - p_z$ in the expression~\eqref{eq:prop_n} at $t < \, t^{\,\prime}$, the $\pm$ sign at $t > \, t^{\,\prime}$ and $t < \, t^{\,\prime}$ still remains just in the sign at $E_n$.
It is appropriate to use the following relation, where the expression for energy~\eqref{eq:E_n} is taken into account:
\begin{equation}
\frac{f(\pm E_n)}{2 \, E_n}\,\ee^{\mp \, \ii \, E_n (t - t^{\,\prime})}
\Bigr|_{t \gtrless \, t^{\,\prime}} = \frac{\ii}{2 \pi} 
\int_{-\infty}^{+\infty} \frac{\dd p_0 \, f(p_0)\,
\ee^{- \, \ii \, p_0 (t - t^{\,\prime})}}
{p_{\parallel}^2 - m^2 - 2 \, \beta \, n + \ii\,\varepsilon}\,,
\label{eq:relation}
\end{equation}
where $p_{\parallel}^2 = p_0^2 - p_z^2$. Hereafter we use the indices ``$\parallel$'' and ``$\perp$'' for denoting the $4\mbox{-}$vector components, belonging to the pseudo-Euclidean subspace $(0, \,z)$ and the Euclidean plane $(x, \, y)$: 
$(a \, b)_{\parallel} = a_0 \, b_0 - a_z \, b_z$, 
$(a \, b)_{\perp} = a_x \, b_x + a_y \, b_y$, 
$(a \, b) = (a \, b)_{\parallel} - (a \, b)_{\perp}$. 

Using the relation~\eqref{eq:relation} we add to the expression~\eqref{eq:prop_n} an integration over the zero momentum component. As a~result the propagator can be written at $t > \, t^{\,\prime}$ and at $t < \, t^{\,\prime}$ identically. Renaming the variables $\xi^{(+)} = \xi$, $\xi^{(+) \,\prime} = \xi^{\,\prime}$, we reduce~\eqref{eq:prop_n} with taking into account~\eqref{eq:preob} and~\eqref{eq:relation} to the form of:
\begin{eqnarray}
&& S_n (X, X^{\,\prime}) = \frac{\ii}{2^n \, n!} \, \sqrt{\frac{\beta}{\pi}} \, 
\exp \left(- \, \beta \, \frac{x^2 + x^{\,\prime\,2}}{2} \right) 
\int\frac{\dd p_0 \, \dd p_y \, \dd p_z}{(2 \pi)^3 }\times 
\nonumber\\[3mm]
&&\times 
\frac{\ee^{- \, \ii \, \left( p \,(X - X^{\,\prime}) \right)_{\parallel}}}
{p_{\parallel}^2 - m^2 - 2 \, \beta \, n + \ii\,\varepsilon}\,
\exp \left\{ - \, \frac{p_y^2}{\beta} 
- p_y \left[\,x + x^{\,\prime} - \ii \, (y - y^{\,\prime})\right] \right\}\times 
\nonumber\\[3mm]
&&\times 
\bigg\{ 
\left[ (p \gamma)_{\parallel} + m \right]
\left[ \varPi_- \, H_n (\xi) \, H_n (\xi^{\,\prime}) 
+ \varPi_+ \, 2 n \, H_{n-1} (\xi) \, H_{n-1} (\xi^{\,\prime}) 
\right] + 
\nonumber\\[3mm] 
&& + \,
\ii \, 2 n \, \sqrt{\beta} \, \gamma^1 
\left[ \varPi_- \, H_{n-1} (\xi) \, H_n (\xi^{\,\prime}) 
- \varPi_+ \, H_n (\xi) \, H_{n-1} (\xi^{\,\prime}) 
\right]
\bigg\} \, .
\label{eq:prop_n2}
\end{eqnarray}

It is worthwhile to note that the expression~\eqref{eq:propagator_sum_n} with~\eqref{eq:prop_n2} for 
the~electron propagator in a~constant uniform magnetic field as the~sum over Landau levels in the 
$x\mbox{-}$space has its own significance. In some cases, this form of the propagator can be more 
convenient than other representations. 

One can make an integration over~$p_y$ in the propagator~\eqref{eq:prop_n2} by introducing a~new variable
$$u = \frac{p_y}{\sqrt{\beta}} 
+ \frac{\sqrt{\beta}}{2} \left[\,x + x^{\,\prime} - \ii \, (y - y^{\,\prime})\right],$$
and using the well-known integrals:\cite{Prudnikov}
\begin{eqnarray}
&& \int_{-\infty}^{\infty} \ee^{- u^2} 
\, H_n (u + a) \, H_n (u + b) \, \dd u = 
2^n \, n! \, \sqrt{\pi} \, L_n (- 2\, a \, b) \,,
\nonumber\\
&& \int_{-\infty}^{\infty} \ee^{- u^2} 
\, H_n (u + a) \, H_{n-1} (u + b) \, \dd u = 
\nonumber\\
&& = 2^{n-1} \, n! \, \sqrt{\pi} \, \frac{1}{b} \,
\left[ L_n (- 2\, a \, b) - L_{n-1} (- 2\, a \, b)\right] ,
\label{eq:interg}
\end{eqnarray}
where $L_n (x)$ are the Laguerre polynomials:
\begin{equation}
L_n (x) = \frac{1}{n!} \, \ee^{x} \, \frac{\dd^n}{\dd x^n} 
\left( x^n \, \ee^{-\,x} \right)\, .
\label{eq:ChebLag}
\end{equation}

As a~result, the $n$th Landau level contribution into the electron propagator in a magnetic field can be presented in the form:
\begin{equation}
S_n (X,X^{\,\prime}) = \ee^{\ii \, \varPhi(X,\,X^{\,\prime})} \, 
{\Hat S}_n (X - X^{\,\prime}) \,,
\label{eq:prop_n3}
\end{equation}
where $\varPhi(X,X^{\,\prime})$ is the translational and gauge non-invariant phase, which is equal for 
all Landau levels: 
$$\varPhi(X,X^{\,\prime}) = -\,\frac{\beta}{2} \, 
(x + x^{\,\prime})(y - y^{\,\prime})\,.$$
For more details about properties of the phase, see, e.\,g., Ref.~\refcite{Kuznetsov:2003}.
${\Hat S}_n (Z)$ is the gauge and translational invariant part of the propagator ($Z = X - X^{\,\prime}$), represented in the form of the double integral over $p_{\parallel}$: 
\begin{eqnarray}
{\Hat S}_n (Z) 
&=& \frac{\ii\,\beta}{2\pi}\, \exp \left( -\,\frac{\beta}{4}\,Z_{\bot}^2 \right)
\int \frac{\dd^2 p_{\parallel}}{(2\pi)^2}\,
\frac{\ee^{-\,\ii \, (p\,Z)_{\parallel}}}
{p_{\parallel}^2 - m^2 - 2 \, \beta \, n + \ii\,\varepsilon}\times
\nonumber\\[3mm]
&\times& 
\left\{ 
\vphantom{\frac{(X\,\gamma)_{\bot}}{X_{\bot}^2}}
\Bigl[(p\,\gamma)_{\parallel} + m \Bigr] 
\left[ \varPi_{-} \, L_n \left( \frac{\beta}{2}\,Z_{\bot}^2 \right) + 
\varPi_{+} \, L_{n-1} \left( \frac{\beta}{2}\,Z_{\bot}^2 \right) \right] +
\right. 
\nonumber\\[3mm]
&+& \left.
2\,\ii\,n\,\frac{(Z\,\gamma)_{\bot}}{Z_{\bot}^2} \,
\left[ L_n \left( \frac{\beta}{2}\,Z_{\bot}^2 \right) 
- L_{n-1} \left( \frac{\beta}{2}\,Z_{\bot}^2 \right) \right]
\right\}.
\label{eq:prop_n4}
\end{eqnarray}
%

\section{The Electron Propagator in a Magnetic Field\\ as the~Sum over Landau Levels}
\label{sec:electron-propagator-as-sum}

Let us compare the obtained expression~\eqref{eq:prop_n4} with the available expansion of the propagator over Landau levels, given in Ref.~\refcite{Chodos:1990}. In that paper the propagator had the form analogous to~\eqref{eq:prop_n3}, but its gauge and translational invariant part was presented in the form of integral 
over $4\mbox{-}$momentum (see note added in proof):
\begin{equation}
{\Hat S} (Z) = \int \frac{\dd^4 p}{(2\pi)^4} \,
S (p) \, \ee^{-\,\ii \, (p\,Z)} \,, \qquad
S(p) = \sum\limits_{n=0}^{\infty} S_n (p) \,, 
\label{eq:prop_Chodos1}
\end{equation}
where
\begin{eqnarray}
%
S_n (p) &=& \frac{\ii}{p_{\parallel}^2 - m^2 - 2 \, \beta \, n 
+ \ii\,\varepsilon} 
\biggl\{\Bigl[(p\,\gamma)_{\parallel} + m\Bigr]
\left[d_n(v) - \frac{\ii}{2}\,(\gamma\,\varphi\,\gamma)\,d_n^{\,\prime}(v)\right] -
\nonumber\\[3mm]
&-& 2\,n\,\frac{d_n(v)}{v}\,(p\,\gamma)_{\bot}\biggr\} \, ,
\quad \quad v = \frac{p_\bot^2}{\beta} \,.
\label{eq:prop_Chodos2}
\end{eqnarray}
The functions $d_n(v)$ have the form:
\begin{equation}
d_n(v) = (-1)^n\,\ee^{-v} \, \bigl[L_n(2\,v) - L_{n-1}(2\,v)\bigr] \,,
\label{eq:d_n}
\end{equation}
where $L_n (x)$ are the Laguerre polynomials~\eqref{eq:ChebLag} with an additional definition $L_{-1} (x) \equiv 0$. The expression $(\gamma\,\varphi\,\gamma) = \gamma_\alpha\,\varphi^{\alpha \beta}\,\gamma_\beta$ contains the dimensionless tensor of the external magnetic field $\varphi^{\alpha \beta} = F^{\alpha \beta}/B$. In the frame, where the field is directed along the $z$ axis, one has $(\gamma\,\varphi\,\gamma) = - 2 \,\gamma^1\,\gamma^2$.

To ensure that our expression for the propagator~\eqref{eq:prop_n4} is consistent with  Eqs.~\eqref{eq:prop_Chodos1},~\eqref{eq:prop_Chodos2}, it is enough to perform in Eq.~\eqref{eq:prop_Chodos1} the integration over the momentum components $p_x$, $p_y$, which are transverse to the field. Thus, the $n$th Landau level contribution to the propagator is expressed via three different integrals $\mathrm{I}_{1,\,2,\,3} (Z_\bot)$ in the Euclidean plane ($p_x, \, p_y$):
\begin{eqnarray}
&&{\Hat S}_n (Z) = \int\frac{\dd^2 p_{\parallel}}{(2\pi)^2}\,
\frac{\ii\,\ee^{-\,\ii\,(p\,Z)_\parallel}}{p_{\parallel}^2 - m^2 
- 2 \, \beta \, n + \ii\,\varepsilon} \times
\nonumber\\[3mm]
&&\times \left\{
\Bigl[(p\,\gamma)_\parallel + m \Bigr] \left[\mathrm{I}_1 (Z_\bot) 
- \frac{\ii}{2}\,
(\gamma\,\varphi\,\gamma)\,\mathrm{I}_2 (Z_\bot)\right] 
- 2\,n\,\mathrm{I}_3 (Z_\bot)
\right\}.
\label{eq:prop_Chodos3}
\end{eqnarray}
An integration over the polar angle leads to the Bessel integral:
\begin{equation}
\int_{0}^{2\pi} \ee^{\ii\,(\xi\,\cos\varphi - n \, \varphi)} \, 
\dd\varphi = 2\,\pi\, \ii^n J_n(\xi) \,,
\label{eq:Bessel}
\end{equation}
where $J_n(\xi)$ is the Bessel function. As a~result, the integrals $\mathrm{I}_{1,\,2,\,3} (Z_\bot)$ take the form:
\begin{eqnarray}
\mathrm{I}_1 (Z_\bot) &=& \int\frac{\dd^2 p_\bot}{(2\pi)^2}\,
d_n(v) \, \ee^{\ii\,(p\,Z)_\bot} =
\frac{\beta}{4 \pi}\,\int\limits_{0}^{\infty}\dd v\,
J_0(\sqrt{\beta}\,Z_\bot\,\sqrt{v})\,d_n(v) \,,
\nonumber\\[3mm]
\mathrm{I}_2 (Z_\bot) &=& \int\frac{\dd^2 p_\bot}{(2\pi)^2}\,
d_n^{\,\prime}(v)\,\ee^{\ii\,(p\,Z)_\bot} =
\frac{\beta}{4 \pi}\,\int\limits_{0}^{\infty}\dd v\,
J_0(\sqrt{\beta}\,Z_\bot\,\sqrt{v})\,d_n^{\,\prime}(v) \,,
\nonumber\\[3mm]
\mathrm{I}_3 (Z_\bot) &=& \int\frac{\dd^2 p_\bot}{(2\pi)^2}\,
\frac{d_n(v)}{v}\,\ee^{\ii\,(p\,Z)_\bot} \, (p\,\gamma)_\bot = 
\nonumber\\[3mm]
&=& \ii\,\frac{\beta^{3/2}}{4 \pi}\,\frac{(Z\,\gamma)_\bot}{Z_\bot}
\,\int\limits_{0}^{\infty}\dd v\,
J_1(\sqrt{\beta}\,Z_\bot\,\sqrt{v})\,\frac{d_n(v)}{\sqrt{v}}\,,
\nonumber
\end{eqnarray}
where $Z_\bot = \sqrt{Z_\bot^2} = \sqrt{(x - x^{\,\prime})^2 + (y - y^{\,\prime})^2}$. Calculating the  integrals:\cite{Prudnikov}

\begin{eqnarray}
%
\mathrm{I}_1 (Z_\bot) &=& \frac{\beta}{4\pi} \, \exp \left( -\,\frac{\beta}{4}\,Z_{\bot}^2 \right)
\left[L_n \left( \frac{\beta}{2}\,Z_{\bot}^2 \right) + 
L_{n-1} \left( \frac{\beta}{2}\,Z_{\bot}^2 \right)\right],
\nonumber\\[5mm]
\mathrm{I}_2 (Z_\bot) &=& -\,\frac{\beta}{4\pi} \,
\exp \left( -\,\frac{\beta}{4}\,Z_{\bot}^2 \right) 
\left[L_n\left( \frac{\beta}{2}\,Z_{\bot}^2 \right) 
- L_{n-1}\left( \frac{\beta}{2}\,Z_{\bot}^2 \right)\right],
\nonumber\\[5mm]
\mathrm{I}_3 (Z_\bot) &=& -\,\ii\,\frac{\beta}{2\pi}\,
\frac{(Z\,\gamma)_\bot}{Z_\bot^2}
\,\exp \left( -\,\frac{\beta}{4}\,Z_{\bot}^2 \right) \, 
\left[L_n\left( \frac{\beta}{2}\,Z_{\bot}^2 \right) 
- L_{n-1}\left( \frac{\beta}{2}\,Z_{\bot}^2 \right)\right],
\nonumber
\end{eqnarray}
and substituting them into~\eqref{eq:prop_Chodos3}, one finally obtains the expression, which coincides with~\eqref{eq:prop_n4}.

\section{Conclusion}
\label{sec:Conclusion}

We have constructed the $n$th Landau level contribution to the exact electron propagator in an~external magnetic field, based on the Dirac equation exact solutions. It is performed in the $x\mbox{-}$representation. 
The expression~\eqref{eq:propagator_sum_n} with~\eqref{eq:prop_n2} for the propagator can be more 
convenient than other representations in some cases. 
This result could have a methodological significance for further developments of the calculation technique for the analysis of quantum processes in an external active medium such as hot dense plasma and strong electromagnetic fields. In particular, it could be useful in the situation of the moderately large field strengths, when it is insufficient to take into account only the ground Landau level contribution. 


\section*{Note added in proof}
There was an error in expression for the propagator in Ref.~\refcite{Chodos:1990}, 
namely, the term in the second line of Eq.~(4.33) should contain the factor $(-i)$. This error was corrected in Ref.~\refcite{Chyi:2000}, Eqs.~(39) and (40), 
and also in Ref.~\refcite{Gusynin:1999}, Eqs.~(13) and (14), 
but without any comments. We thank M.\,I.~Vysotsky for a discussion clarifying this point.

\section*{Acknowledgments}

We are grateful to N.\,V.~Mikheev, M.\,V.~Chistyakov and D.\,A.~Rumyantsev for useful remarks.

This work was performed in the framework of realization of the Federal
Target Program ``Scientific and Pedagogic Personnel of the Innovation
Russia'' for 2009\,--\,2013 (State contract no. P2323)
and was supported in part by the Ministry of Education
and Science of the Russian Federation under the Program
``Development of the Scientific Potential of the Higher
Education'' (project no. 2.1.1/13011), and by the Russian Foundation
for Basic Research (project no. 11-02-00394-a).



\end{document}